# PHYSICAL ASPECTS OF CANCER INVASION

*Caterina Guiot* [1,2], *Nicola Pugno* [3], *Pier Paolo Delsanto* [2,4,5], *Thomas S. Deisboeck* [6,*]

[1] Dept. Neuroscience, University of Torino, Torino, Italy;

[2] CNISM, Torino, Italy;

[3] Dept. of Structural Engineering and Geotechnics, Politecnico of Torino, Torino, Italy;

[4] Dept. of Physics, Politecnico of Torino, Torino, Italy;

[5] Bioinformatics and High Performance Computing Labs, Bioindustry Park of Canavese, Colleretto Giacosa, Italy;

[6] Complex Biosystems Modeling Laboratory, Harvard-MIT (HST) Athinoula A. Martinos Center for Biomedical Imaging, Massachusetts General Hospital, Charlestown, MA 02129, USA.

**\*Corresponding author:**

Thomas S. Deisboeck, M.D.

Complex Biosystems Modeling Laboratory

Harvard-MIT (HST) Athinoula A. Martinos Center for Biomedical Imaging

Massachusetts General Hospital-East, 2301

Bldg. 149, 13th Street

Charlestown, MA 02129

Tel: 617-724-1845

Fax: 617-726-7422

Email: deisboec@helix.mgh.harvard.edu

**Abstract**

Invasiveness, one of the hallmarks of tumor progression, represents the tumor's ability to expand into the host tissue by means of several complex biochemical and biomechanical processes. Since certain aspects of the problem present a striking resemblance with well known physical mechanisms, such as the mechanical insertion of a solid inclusion in an elastic material specimen [1, 2] or a water drop impinging on a surface [3], we propose here an analogy between these physical processes and a cancer system's invasive branching into the surrounding tissue. Accounting for its solid and viscous properties, we present a unifying concept that the tumor behaves as a *granular solid*. While our model has been explicitly formulated for multicellular tumor spheroids *in vitro*, it should also contribute to a better understanding of tumor invasion *in vivo*.

## 1. Introduction

Due to the complexity of the mechanisms involved in neoplastic growth, *in silico* experiments are becoming the tool of choice [4] for the description and interpretation of the observed phenomenology, based on models ranging from microscopic to macroscopic. *Microscopic* models are suitable to describe malignant transformations, which lead to profound alterations at the sub-cellular and cellular levels. *Macroscopic* models of solid tumor growth may be related to universal scaling laws [5], in analogy with the model of ontogenic growth for all living organisms, proposed by G. B. West and collaborators [6,7]. A similar quest for universalities is currently being pursued in a completely different scientific context (continuum mechanics and elasticity) [8] and extended to other fields as well [9]. In all cases, a bridging between a microscopic and a macroscopic description is of fundamental importance [10] and can be best achieved by means of intermediate, so called *mesoscopic* models [11-14].

An abundance of experimental results indicates that the process of tissue invasion [15] depends not only on the characteristics of the malignant cells, but also on the surrounding microenvironment or, more generally, on the properties of the host tissue, including its status of aging and any other changes or damages that alter its condition [16-18]. On the other hand, the matrix' reaction to the pressure exerted by the growing (solid) tumor and its effects of cells compaction, compression and stromal degradation at its boundaries may be well described macroscopically.

Several theoretical models of tumor invasion have been proposed, either describing the tumor-host interface as a traveling wave edge [19], through extracellular matrix degradation [20], or for specific applications to invasive gliomas [21]. The role of tensional homeostasis on malignant growth has also been investigated at a microscopic level [22]. Recent observations of invasive branching (or fingering) in multicellular tumor spheroids (MTS) [23, 24] suggest that invasion cannot be solely caused by tumor cell proliferation, rather, that growth instabilities are also required at the interface with the host. In fact, a remarkable amount of information about the biochemical mechanisms occurring at the 'tumor-host' interface is now available. Hence, one can learn how the local host stroma is affected by the degrading enzymes produced by the tumor itself [25], how the extracellular matrix (ECM) is remodelled by endogenous substances (integrins, focal adhesion kinases) [26] and how all these molecular mechanisms cross-interact [27]. No simple model can account for such a detailed and complex experimental background. There are, however, some unifying features that should be noted:

1) *Short-range processes*, i.e. the enzyme cascade is confined to the cell surface of the invading pseudopodia [28].

2) *Borderline processes* have important implications also for nutrients availability: the appearance of sprouts and invasive branches affects the tumor's surface-to-volume ratio. The fractal dimension (and other related parameters, such as the scaling parameter of the 'West-like' growth law [16, 29]) are therefore increased with respect to 2/3, which would correspond to diffusion across a spherical surface. Similar patterns have been observed in bacterial colonies cultured in more rigid media (i.e. a high concentration of agar) and poor nutrient concentration [30].

**2. Why an amorphous 'solid – fluid' model?**

Here, we attempt to reconcile the solid and the fluid mechanical models [2,3], starting from the fact that, in many ways, an amorphous solid can behave as a fluid and viceversa. Thus, fluid- and solid-like behaviours often coexist. For instance, it is well known that the earth crust, during asteroid or meteorite impacts, displays a fluid-like behaviour. In particular, the classical bowl-like shaped crater often shows a central pinnacle, due not simply to elastic recoil or melting, but as a consequence of drop-like reflux [31]. Conversely, a fluid under strong shear stress behaves as a solid, a fact which is used to smartly increase the protective capability of 'liquid' armors, i.e. solid armors impregnated by colloidal suspensions [32]. An attempt to treat in a unified manner solids and liquids is represented by the theory of *granular solids* (or, conversely, of viscous fluids). In fact, such systems behave as an amorphous solid if their grains are confined, but as a viscous fluid if they are not [33]. Accordingly, the model formulated in the present paper represents a first step in the direction of treating a tumor as a granular solid.

Using the new theory of Quantized Fracture Mechanics [34-36] we introduce a characteristic length (missing in classical continuum approaches, such as Elasticity or Fluid Mechanics), i.e. the *fracture quantum*, to describe the granular size. For physical lengths (e.g. cracks) much larger than the granular size, the granular solid behaves as a continuum, and classical fracture mechanics is recovered. On the other hand, for characteristic sizes much smaller than the granular size, a fluid-like behaviour prevails Thus, the model predicts a smooth transition from liquid to solid-like behaviour and thus, taken together, yields a more realistic description in which both behaviours coexist.

## 3. An amorphous solid analogy

Guiot *et al.*: Physical Aspects of Cancer Invasion

In this mechanistic analogy, cracks correspond to the cellular infiltration channels of **Figure 1(c)** and failure is usually assumed to arise, for unnotched specimens, when the stress $\sigma$ reaches the material strength tolerance $\sigma_C$.[37]. In notched specimens, it is not the stress $\sigma$ but the stress-intensity factor $K$, which must reach a critical value $K_C$ for fracture propagation [38]. Thus, $K \equiv \chi\sigma\sqrt{\pi l} = K_C$, i.e. the stress-intensity factor at the tip of a crack of length $l$ loaded by a stress $\sigma$ must be equal to the fracture toughness $K_C$ of the material; $\chi$ is a geometrical factor, e.g., for a crack at the edge of a large medium $\chi \approx 1.12$.

Recently, a more powerful criterion (valid both for small and large values of $l$) for predicting the strength of solids has been derived [39], by simply removing the assumption of continuum crack propagation. Accordingly, the failure stress is estimated as:

$$\sigma_f = \frac{\sigma_C}{\sqrt{1+\frac{\pi\chi^2\sigma_C^2}{K_{IC}^2}l}}, \qquad (1)$$

which represents an asymptotic matching between the two previously discussed solutions. Further details, such as the number of branches during invasion, can be deduced as follows: Let us consider a cylindrical tumor of radius $R$ embedded in a linear elastic matrix. Take $N$ cracks of length $a$, starting at and perpendicularly to the interface, equally spaced and thus with an angular period of $2\pi/N$. According to Fracture Mechanics the stress intensity factor at the tip of each crack is:

$$K_I = P\sqrt{\pi a}\,F\!\left(\frac{a}{R+a}, N\right) \qquad (2),$$

where *P* is the tumor-to-matrix interface pressure and *F* is a known function [40]. Note that $F(0,N) \approx 2.243$ [1], whereas $F(1,N) \approx 2/\sqrt{N}$ [42,43], for large value of *N* (N >10). According to Quantized Fracture Mechanics [34-36], propagation will take place when

$$\sqrt{\langle K_I^2 \rangle_a^{a+q}} = K_{IC} \quad (3),$$

in which ($\langle \bullet \rangle_a^{a+q} \equiv \frac{1}{q}\int_a^{a+q} \bullet \, da$) $K_{IC}$ represents the material fracture toughness and *q* the fracture quantum (related to the microstructure) of the matrix. Note that according to classical Fracture Mechanics $q \to 0$. By introducing Eq. (2) into Eq. (3) and inverting the latter with respect to *N*, we obtain the number of branches during tumor invasion:

$$N_f \approx \frac{4\pi P^2 (a+q/2)}{K_{IC}^2} C,$$

with $\quad C = \frac{\langle aF^2 N/4 \rangle_a^{a+q}}{a+q/2} \quad (4)$

For large cracks it follows that $C \approx 1$. We note that *q* can be fixed imposing the same predictions in the case of small cracks with those derivable according to the *splashing water drop* analogy [38]. Accordingly:

$$q \approx \frac{K_{IC}^2}{2\pi}\sqrt{\frac{R}{\sigma P^3}} \quad (5)$$

where $\sigma$ is the surface tension. Thus, for large cracks, $N_f \approx \frac{4\pi P^2 a}{K_{IC}^2}$, whereas for small cracks

$$N_f \approx \sqrt{\frac{PR}{\sigma}} \qquad (6).$$

**4. A viscous fluid analogy**

In 1898, the naval engineer H.J.S. Hele-Shaw observed that a drop of liquid injected in a more viscous environment would generate an instability that leads to a variable number of 'fingers'. The macroscopic details of this so-called 'Hele-Shaw' effect depend on the combination of selected fluids and their viscosity. Recently, this effect has been widely studied because of its intrinsic fractality, and the fractal dimensions have been measured for many pairs of fluids. The appearance of the same patterns in MTS is probably related to the strong viscosity of the commonly used ECM gel MATRIGEL ™ (in the order of 10 Pa s, see www.tbmc.it), while the viscosity of the MTS' is unfortunately unavailable (but probably lower than in Matrigel). Also other common culture media, such as collagen, edible gelatine and agar can reach large viscosity values, up to 100 Pa s, after sol-gel transition.

Apparently, cell membrane viscosity can vary over a wide range of values. For instance, Jiang Yu-Qiang et al. (2004) [43] found for breast cancer cells a very large value (0.021 pN.s/µm$^3$, corresponding to 2.1 10$^4$ Pa s), while Dunham et al. (1996) [44] obtained for keratinocytes values between 60 to 120 cP (i.e. 0.06-0.12 Pa s). Further investigation in biological tissues is rather cumbersome, due to the need of accurate measurements of their viscosity, but in principle it should be possible to predict whether a particular tumor type in a given tissue or organ would exhibit a finger-like invasive pattern or not.

As observed in [3], there is a striking analogy between MTS invasion and a liquid drop impacting on a solid surface and causing the formation of a fluid 'crown' ('Rayleigh' or 'Yarin-Weiss' capillary instability [45-47]) as shown in **Figure 2**. They seem to share several features, although they may not be easily recognized due to the unfamiliar terminology: e.g., the occurrence of invasive 'fingering' corresponds to the secondary jets, the evidence for branch-confluence corresponds to hole nucleation near the fluid rim and, finally, the proliferating aggregates emerging within the invasive cell population [48] correspond to the outgoing small drops at the fluid-air interface. Intriguingly, the number of fingers can then be predicted on the basis of the following parameters: the fluid density $\rho$, the drop radius R, the deceleration a, the fluid viscosity $\mu$ and the surface tension $\sigma$ [49]:

$$N_f = 2 \pi R / \lambda, \qquad (7)$$

where

$$\lambda = 2 \pi ( 3 \sigma / \rho a )^{1/2} \qquad (8)$$

Assuming for simplicity a spherical shape and a radius R at the invasion time:

$$a = F/m = P S / \rho V = 3 P / \rho R, \qquad (9)$$

which, remarkably, yields again Eq. (6).

**5. Discussion**

Following Eq. (6), the value $N_f = 1$ separates the case of no branching (hence no tumor invasion) from the one in which invasion takes place. By defining the dimensionless *Invasion Parameter*,

$$IP = PR/\sigma, \qquad (10)$$

invasive behaviour is to be expected in all but for the case of $IP < 1$ (which implies large tumor surface tension, small confining pressure and/or small tumor radius values). According to Eq. (10), which can be derived with both the 'mechanistic' and the 'viscous fluid' model, tumor invasion is controlled by three parameters, namely:

a) *Tumor surface tension:*

Multicellular aggregates of three different malignant astrocytoma cell lines (U-87MG, LN-229 and U-118MG) investigated by Winters et al. [50] with surface tensions of about 7, 10 and 16 dyne/cm respectively, showed a significant *inverse* correlation between invasiveness and surface tension. Moreover, the anti-invasive therapeutic agent Dexamethason is known to increase the tumor surface tension or cohesivity between cells. Therefore, surface tension can indeed be a predictor of *in vitro* invasiveness, with a threshold value for $\sigma$ of about 10 dyne/cm.

b) *Microenvironmental pressure:*

The mechanical interaction between matrix and tumor has been recently investigated more thoroughly. For instance, Paszek et al. [51] observed that stiffer tissues promote malignant behavior. Similarly, the experimental work on NIH3T3 fibroblasts by Georges and Janmey [52] shows that they keep a roughly spherical shape (suggesting prevalence of cohesive forces) when embedded in a soft polyacrilamide gel, while in a stiff gel they exhibit finger-like patterns (consistent with the preponderance of adhesive forces). We note that the effect of external pressure on the growth of tumor cell colonies has also been studied by Bru and Casero (2006) [53], showing that geometrical and dynamical patterns are markedly dependent on the pressure exerted by the surrounding medium. Moreover, pressure can act either as an inhibitor or as an enhancer for tumor cell proliferation, depending on the particular cell line (see e.g. DiResta et al., 2005 [54]).

c) *Tumor radius*:

Finally, Tamaki et al. [55] investigated C6 astrocytoma spheroids with different diameters (i.e., 370, 535 and 855 μm on average), which were implanted in collagen type I gels. The authors showed that spheroid size indeed correlated with a larger total invasion distance and increased rate of invasion.

As such, any therapeutic strategies that solely or in combination are geared towards *reducing* tumor burden, *diminishing* surrounding mechanical pressure and *increasing* (residual) tumor surface tension, may eventually hold promise in clinics.

**6. Conclusions**

Tumor invasion involves a variety of biochemical and biomechanical processes. In this contribution we have reviewed two models, based on analogies with well known mechanisms of fracture mechanics and fluid dynamics, which have been recently proposed to illustrate some of the features of tumor invasion. In the former, tumors are visualized as amorphous solids, while in the latter as viscous fluids. We have attempted, in this paper, to reconcile the two representations by means of an intermediate one, i.e., the "granular" solid model. Remarkably, the most useful result that we have obtained, i.e. the formula for the so-called *Invasion Parameter*, is consistent with both models (and, of course, with the intermediate one). Such a formula identifies the most relevant physical parameters, whose control should be the target of dedicated therapies, e.g. the tumor's surface tension, its radius and the confining host tissue pressure. Understanding their role could explain why some therapies fail while others prove to be effective in locally controlling tumor expansion. While a reliable, patient-specific assessment of tissue properties poses a formidable challenge, in principle, one should be able to predict whether a particular tumor type in a given host organ exhibits finger-like invasion patterns or not. Eventually, a cancer type, organ site, and patient specific *IP* may be of significant value for diagnostic purposes as most of this multicellular behaviour occurs well below the current non-invasive imaging resolution limits. However, for any such future iteration a more realistic description should obviously take into account the *heterogeneity* of both tumor and micronenvironment, which would not only imply *regional* differences in the IP value but also argue for a *dynamic* behavior of IP. Similarly, the schematic 'sequence' suggested in **Figure 2** would then refer to a single site rather than to the entire tumor, as invasive branching would occur *desynchronized* at numerous sites across the tumor surface.

To conclude, we believe that, although an all-comprehensive model of cancer invasion (or, in general, tumor growth) is desirable, for the time being it may be more expedient to use 'composite' models, in which the different facets of the problem are considered *individually*, not as alternative, but as complementary descriptions. Likewise, for the numerical simulation of neoplastic growth (both for diagnostic and therapeutic purposes), a multilevel approach [56] may be the most promising one that includes both micro- and macroscopic scales and its mesoscopic bridging level.

**Acknowledgements**

This work has been supported in part by NIH grant CA 113004 (Center for the Development of a Virtual Tumor, CViT (http://www.cvit.org)). Helpful discussions with M. Griffa are gratefully acknowledged.

**Figure Captions**

**Figure 1.** Schematic representation of an invasion sequence or 'cycle' (clockwise, from *top left*). **(a)** Initial condition: tumor (*black*) in an elastic matrix (*grey*). **(b)** Non-invasive phase: the interfacial stress increases in the course of the tumor's elastic growth and interaction with the matrix. **(c)** Invasive phase: when the tumor induced stress reaches the matrix strength tolerance threshold, invasion takes place 'ideally' reducing the confining stress to zero (due to matrix-degrading enzymes for instance [57]). **(d)** Final condition: the invasion cycle is concluded and the non-invasive growth phase starts anew. (See text, and [24], for more details).

**Figure 2.** Examples (*top* to *bottom*). **(a)** MTS Tumor [Image reprinted from Habib et al. [58], with permission]. **(b)** Water drop [Image courtesy of Prof. A. Davidazy, Imaging and Photographic Technology, Rochester Institute of Technology, Rochester, NY, URL: http://www.rit.edu/~andpph/exhibit-splashes.html]. **(c)** Scanning Electron Microscopy image of a specimen fractured at an applied stress amplitude of 700 Mpa [Image courtesy of Dr. Z.G. Yang, Shenyang National Laboratory for Material Science, Chinese Academy of Science].

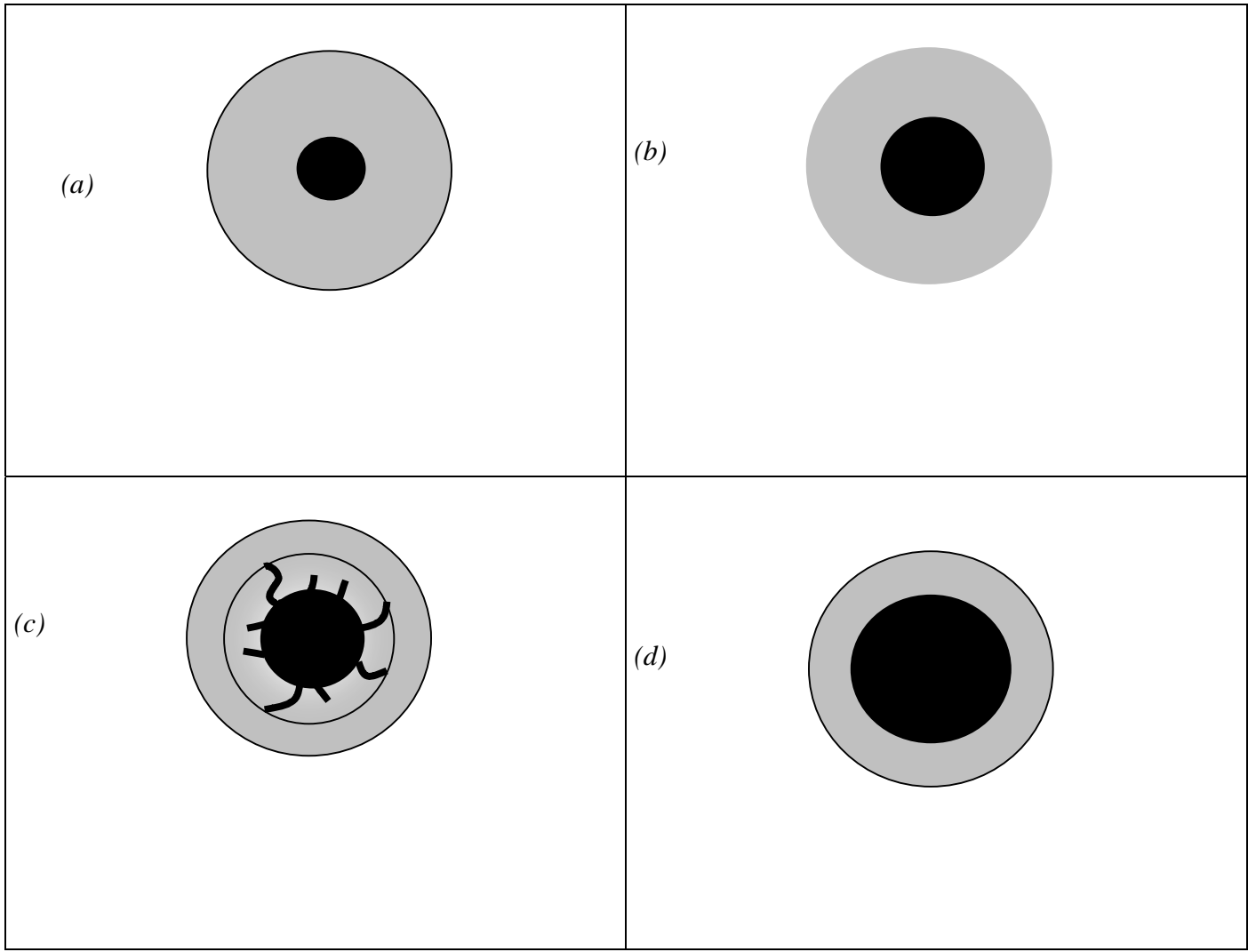

**Figure 1**

Guiot *et al.*: Physical Aspects of Cancer Invasion

**Figure 2**

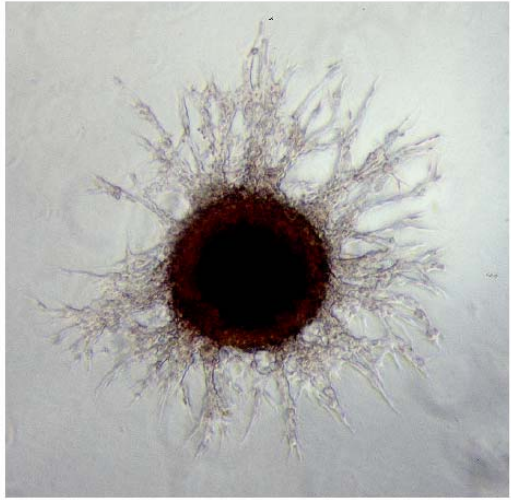

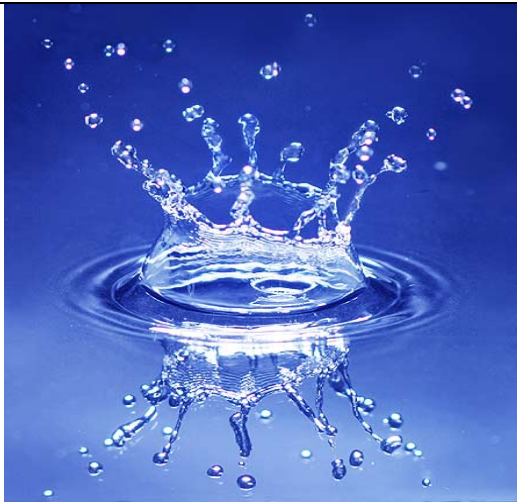

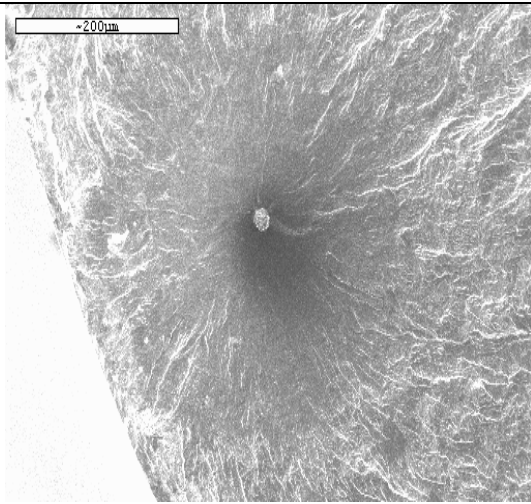